\documentclass[12pt,preprint]{emulateapj}
\slugcomment{{\sc Accepted to ApJL:} February 28, 2013} 

\shortauthors{YAMADA ET AL.}
\shorttitle{HIGHLY IONIZED FE-K ABSORPTION LINE FROM CYGNUS X-1}

\begin{document} 
\title{HIGHLY IONIZED FE-K ABSORPTION LINE FROM CYGNUS X-1 \\ IN THE  HIGH/SOFT STATE OBSERVED WITH SUZAKU
}


\author{S. Yamada\altaffilmark{1}, 
S. Torii\altaffilmark{2}, 
S. Mineshige\altaffilmark{3}, 
Y. Ueda\altaffilmark{4}, 
A. Kubota\altaffilmark{7},
P. Gandhi\altaffilmark{5},\\ 
C. Done\altaffilmark{6}, 
H. Noda\altaffilmark{2},
A. Yoshikawa\altaffilmark{1},
and K. Makishima\altaffilmark{2,1}
}

\altaffiltext{1}{Cosmic Radiation Laboratory,
 Institute of Physical and Chemical Research (RIKEN),
   Wako, Saitama, 351-0198, Japan}
\altaffiltext{2}{Department of Physics, University of Tokyo, 7-3-1 Hongo, Bunkyo-ku, Tokyo, 113-0033, Japan}
\altaffiltext{3}{Department of Astronomy, Kyoto University,
Kitashirakawa-Oiwake-cho, Sakyo-ku, Kyoto 606-8502, Japan}
\altaffiltext{4}{Department of Physics, Kyoto University,
Kitashirakawa-Oiwake-cho, Sakyo-ku, Kyoto 606-8502, Japan}
\altaffiltext{5}{Institute of Space and Astronautical Science, JAXA,
   3-1-1 Yoshinodai, Sagamiharas, Kanagawa, Japan 229-8510}
\altaffiltext{6}{Department of Physics, Durham University, South Road, Durham, DH1 3LE, UK}
\altaffiltext{7}{Department of Electronic Information Systems, Shibaura Institute of Technology,
307 Fukasaku, Minuma-ku, Saitama-shi, Saitama, 337-8570, Japan}


\begin{abstract} 

We present observations of a transient He-like Fe K$_{\alpha}$ absorption line in {\it Suzaku} observations 
of the black hole binary Cygnus X-1 on 2011 October 5 near superior conjunction during the high/soft state, 
which enable us to map the full evolution from the start and the end of 
the episodic accretion phenomena or dips for the first time. 

We model the X-ray spectra during the event and trace their evolution. 
The absorption line is rather weak in the first half of the observation, 
but instantly deepens for $\sim$10 ks, 
and weakens thereafter. 
The overall change in equivalent width is a factor of $\sim$ 3, peaking at an orbital phase of $\sim 0.08$. 
This is evidence that the companion stellar wind feeding the black hole is clumpy. 
By analyzing the line with a Voigt profile, 
it is found to be consistent with 
a slightly redshifted Fe$_{\rm{XXV}}$ transition, 
or possibly a mixture of several species less ionized than Fe$_{\rm{XXV}}$. 
The data may be explained by a clump 
located at a distance of $\sim 10^{10-12}$ cm with a density of $\sim 10^{(-13)-(-11)}$ g cm$^{-3}$, 
which accretes onto and/or transits the line-of-sight to the black hole, 
causing an instant decrease in the observed degree of the ionization 
and/or an increase in density of the accreting matter. 
Continued monitoring for individual events with future X-ray calorimeter missions 
such as ASTRO-H and AXSIO will allow us to map out the accretion environment in detail and how it changes between the various accretion states.


\end{abstract}


\section{INTRODUCTION}

Mass accretion onto a black hole binary (BHB) produces energetic astrophysical phenomena.
BHBs are largely classified into two distinct states: 
the high/soft and the low/hard state (e.g., Remillard \& McClintock 2006). 
Mass accretion is believed to be provided by the stellar wind from the companion star
for high-mass X-ray binaries, 
while it is caused by Roche lobe overflow for low-mass X-ray binaries. 
X-ray absorption features have been observed from both systems, 
and interpreted as the absorption by the disk wind, 
the stellar wind, or the hot spot on the disk. 
Such an absorption feature can be crucial to 
probe the global structure of accretion flow 
that is presumably related to the differences between two states.  

Cyg X-1 is a famous BHB, 
and belongs to one of the high-mass X-ray binaries (Oda et al. 1971, Tananbaum et al.~1972). 
It has an O9.7 Iab supergiant, HD 226868 (Caballero-Nieves et al.~2009). 
The distance, the mass, and inclination are 
$D=1.86^{+0.12}_{-0.11}$ kpc (Reid et al.~2011, Xiang et al.~2011), 
$M=14.8\pm1.0 M_{\odot}$, and $i = 27.1\pm0.8^{\circ}$ (Orosz et al.~2011), respectively.  
The orbital period is  $5.599829$ days (Brocksopp et al. 1999).  
Since HDE~226868 almost fill its Roche lobe in Cyg X-1 (Gies \& Bolton 1986), 
the stellar wind is not spherically symmetric but strongly enhanced towards the BH (``focused wind''; Friend \& Castor 1982). 

The stellar wind causes episodes of increased absorption, 
or ``dips'', which are seen as a decrease in soft X-ray flux 
 (e.g., Pravdo et al.~1980; Kitamoto et al.~1984)
mostly at near the superior conjunction of the BH 
(when the observer, the companion star, and the BH are lined in this order). 
In long-term continuous monitoring with {\it RXTE}, 
these dips have rarely been seen in the high/soft state 
in contrast to the frequent occurrence in the low/hard state (Wen et al. 1999). 
The detailed studies on the absorption feature
have been realized with {\it Chandra} HETG observation of Cyg X-1 
in several orbital phases $\phi_{\rm{orb}}$ and states; e.g., 
$\phi_{\rm{orb}}\approx0.74$ (Schulz et al. 2002)
and $\phi_{\rm{orb}}\approx0$ (Hanke et al. 2009)
in the low/hard state,  
$\phi_{\rm{orb}}\approx0.77$ (Miller et al. 2005) 
in the intermediate state, 
and $\phi_{\rm{orb}}\approx0.88$ (Chang \& Cui et al. 2007) 
and $\phi_{\rm{orb}}\approx0$ (Feng et al. 2003) 
in the high/soft state. 
Despite these observations, 
the differences in absorbing matter between the low/hard state and high/soft state  
are poorly understood, due probably to fewer observations in the high/soft state. 

Cyg X-1 usually stays in the low/hard state, 
though it recently has changed into the high/soft state since the middle of 2010. 
Thus, we observed Cyg X-1 twice with {\it Suzaku} (Mitsuda et al.~2007) in the high/soft state, 
and found clear absorption features at around the Fe-K band in the latter observation 
performed at $\phi_{\rm{orb}}\approx0.05$. 
In this letter, we report the detection 
of the He-like Fe K$_{\alpha}$ absorption line with {\it Suzaku} and its full evolution. 
Unless otherwise stated, errors refer to 90\% confidence limits.

\begin{figure*}
\begin{center}
\vbox{}
\includegraphics[width=0.98\textwidth]{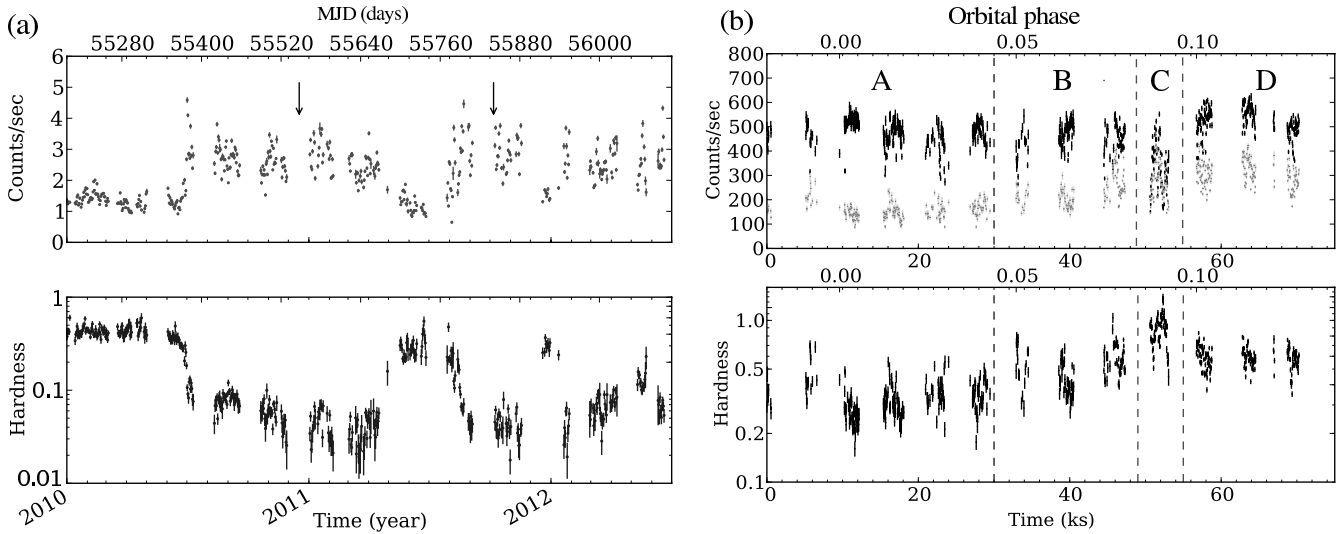}
\caption{
(a) The 2--20 keV Cyg X-1 lightcurve of MAXI GSC, and the hardness of 10--20 keV to 2--4 keV. The two arrows indicate Obs.~1 and Obs.~2, respectively. 
(b) The 0.5--1.5 keV (black) and 3--10 keV (gray) lightcurves and the hardness of 3--10 keV to 0.5--1.5 keV, taken from {\it Suzaku} Obs.~2 with the time bin size of 32s. The orbital phases are plotted on the top. Phase~A, B, C, and D are indicated in the plot.}
\label{fig:lc}
\end{center}
\end{figure*}

\section{OBSERVATION AND ANALYSIS} 

\subsection{Observation} 

Cyg X-1 was observed with {\it Suzaku} on 2010 December 16 (Obs.~1; ObsID=905006010) and 2011 October 5 (Obs.~2; ObsID=406013010). The corresponding orbital phases are 
$\phi_{\rm{orb}} \approx 0.64$--$0.80$ for Obs.~1 and 
$\phi_{\rm{orb}} \approx 0.98$--$0.13$ for Obs.~2, 
assuming that an epoch of the superior conjunction of the BH 
is MJD$41874.207$ based on Brocksopp et al. (1999). 
The XIS employed the 1/4 window option 
and a burst mode of 0.5 s (Koyama et al. 2007; see Serlemitsos et al. 2006 on the X-ray mirror). 
The HXD-PIN and HXD-GSO were operated in the standard mode (Takahashi et al. 2007; Kokubun et al. 2007; Yamada et al. 2011). 

The long-term light curve and hardness of Cyg X-1 
taken from the archival data of MAXI Gas Slit Camera (MAXI: Matsuoka et al.~2009; GSC: Sugizaki et al.~2011) are shown in Figure \ref{fig:lc}a. 
Although Cyg X-1 was believed to stay mostly in the low/hard state, 
it has been in the high/soft state (the hardness ratios $\lesssim 0.2$) over the last two years. 
The first and second observation were performed during the high/soft state 
as indicated in Figure~1a.   

\subsection{Data reduction}

The data processing and reduction were performed 
by using the {\it Suzaku} pipeline processing ver.\,2.7.16.30.
The XIS data are reprocessed by running {\tt xispi} 
with the latest gain calibration file of {\tt ae\_xi?\_makepi\_20120527.fits}. 
Events were discarded if they were acquired in the South Atlantic Anomaly, 
or in regions of low cutoff rigidity 
($\leq 6$ GV for both XIS and HXD), or with low Earth elevation angles. 
The exposures of the XIS and HXD data in Obs.~1 and Obs.~2 were 
2.9~ks and 28.4~ks, and 3.8~ks and 26.3~ks, respectively. 
The reason that the exposure of the XIS is shorter than that of the HXD 
is due to utilizing the burst option (live time fraction of 25\%) 
and discarding the period that the telemetry of the XIS is saturated.
We subtracted the modeled non-X-ray backgrounds (Fukazawa et al. 2009) from the HXD spectra, 
while did not subtract the background from the XIS spectra 
because the background is less than $\sim 0.1$\% in any energy bands below $\sim$ 10 keV. 
We used a standard set of the response files and auxiliary files; i.e., 
ones generated by running {\tt xisrmfgen} and {\tt xissimarfgen} (Ishisaki et al. 2007), 
and provided as the CALDB files for the {\it Suzaku} users. 


We estimated pileup effects on the XISs and excised the image core 
within a pileup fraction larger than 3\% ($\sim 1\arcmin$) 
according to the empirical criteria (Yamada et al. 2012). 
Since the pileup effect distorts the continuum 
but leaves the narrow line features rather untouched, 
we used a circular region of the XIS for Obs.~2 to maximize photon statistics, 
together with an additional auxiliary file to correct for continuum distortion caused by pileup. 
Comparing the XIS spectrum including the image core with the one excluding the core 
when those are used with the auxiliary files, 
we confirmed that its difference is less than $\sim$ 2\%,  
and thus we adopted the systematic uncertainty of 2\% when performing spectral fitting. 

\subsection{Wide-band spectra}

\begin{figure}
\begin{center}
\vbox{}
\includegraphics[width=0.49\textwidth]{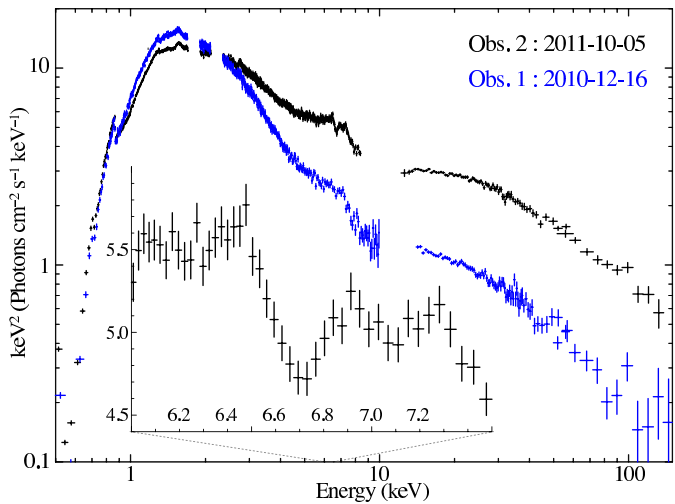}
\caption{Cyg X-1 wide-band spectra in the $\nu F_\nu$ form in Obs.~1 (blue) and Obs.~2 (black) taken from {\it Suzaku} XIS, PIN and GSO. The 6--7.5 keV band of Obs.~2 are inserted in the bottom left panel. The 1.7--1.9 keV and 2.1--2.3 keV band are discarded due to instrumental uncertainties on Si and Au edges.}
\label{fig:avespec}
\end{center}
\end{figure}

Figure 2 shows wide-band spectra taken from Obs.~1 and Obs.~2. 
The spectrum of the former shows a prominent disk emission below $\sim$ 10 keV, 
and a power-law tail with reflected emission, which are typically seen in the high/soft state. 
On the other hand, the spectrum of the latter shows not only more enhanced powerlaw emission, 
but also absorption features at $\sim 6.7$~keV and $\sim$ 7.1 keV as seen in a subset in Figure 2. 

We perform in this letter the analysis of the XIS0 data in Obs.~2 
to concentrate on the absorption features. 
Note that 
such an absorption feature can be seen in the XIS3 and XIS1 spectra, 
but the XIS3 is operated on timing mode, and the XIS1 is a backside illuminated CCD,  
which make it hard to treat all the XISs in a comprehensive manner, 
especially under conditions being affected by pileup. 
More detailed analysis on the wide-band spectra will be described in a subsequent paper. 

\subsection{Lightcurves}

Figure~1b shows the light curves of the XIS0 in Obs.~2. 
The 0.5--1.5 keV count rate is approximately constant at $\sim$ 500 c s$^{-1}$,  
though it significantly decreases to $\sim$ 300 c s$^{-1}$ at $\phi_{\rm{orb}} \simeq 0.08$. 
Meanwhile, the 3--10 keV count rate is rather monotonically increasing 
from $\sim$ 100 c s$^{-1}$ to $\sim$ 300 c s$^{-1}$ over the entire observation, 
and does not show significant decrease at $\phi_{\rm{orb}} \simeq 0.08$. 

The hardness ratio, as shown in the bottom of Figure~1b, 
has gradually increased to $\sim$ 1.0 towards $\phi_{\rm{orb}} \approx 0.08$, 
and afterward decreases to $\sim$ 0.5. 
This rapid change in hardness around $\phi_{\rm{orb}} \approx 0.08$ 
is probably caused by dips, while the gradual spectral changes over the entire observation 
seem intrinsic to the central source.  
To study more concretely spectral evolution depending on the luminosity and $\phi_{\rm{orb}}$, 
we define Phase A, B, C and D, as shown in Figure 1b, to examine each spectrum in detail. 
The dip marginally appears in Phase~A and Phase~B, 
and certainly occurs in Phase~C, and gradually disappears in Phase~D.

\begin{figure}
\begin{center}
\vbox{}
\includegraphics[width=0.46\textwidth]{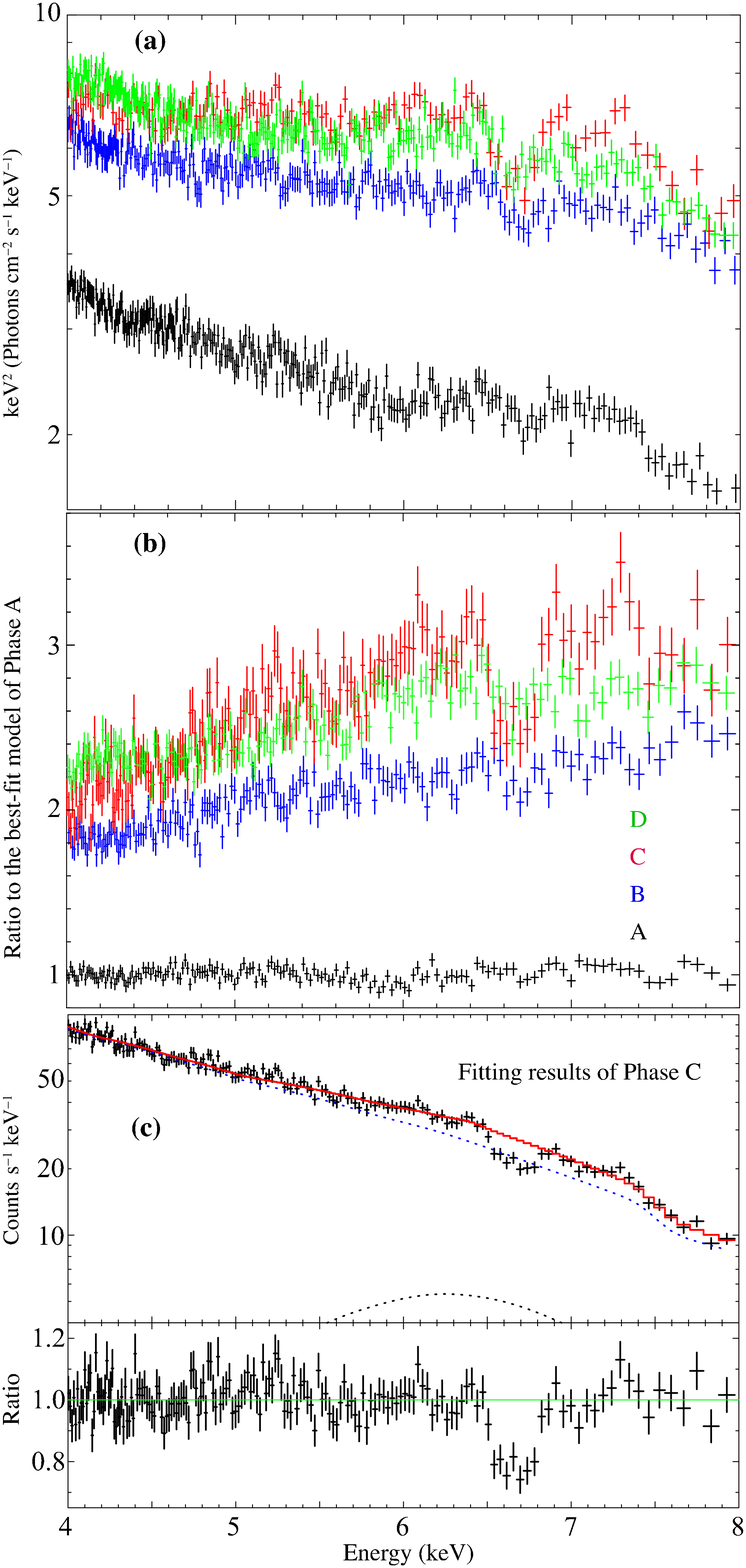}
\caption{
(a) Cyg X-1 $\nu F \nu$ spectra of XIS0 for the four phases. 
(b) The spectral ratios of the four spectra to the best-fit model of Phase A. 
The colors of the labels indicate each phase. 
(c) The best-fit model shown in red and the detector-response-convolved spectrum of Phase C, 
and the spectral ratio to the model except for the absorption line. 
}
\label{fig:specfit}
\end{center}
\end{figure}

\section{SPECTRAL ANALYSIS AND RESULTS} 

\subsection{Spectral features in each phase} 

Figure 3a shows unfolded spectra of XIS0 taken from the four phases.  
The spectrum becomes harder as the source becomes brighter by a factor of $\sim 3$. 
The deepest absorption feature at $\sim$ 6.7 keV can be most clearly seen in Phase C. 
In contrast, the absorption feature in Phase A is shallower than the others. 
The overall shape of the spectra is not straight even in such a narrow band, 
which is caused by a broad Fe-K emission line, and Fe edges (Dotani et al.~1997), 
and concave continua probably consisting of blend of disk emission, non-thermal emission, 
and associated reflection. The similar trend is reported in the previous study (e.g., Feng et al. 2003).

Since our aim is to quantify absorption lines and temporal variations,  
we employed conventional models implemented in {\tt Xspec} 
as an underlying continuum to fit the spectra; 
i.e., {\tt tbabs*edge*(diskbb+power)}, 
where an interstellar absorption {\tt tbabs} (Wilms et al. 2000), 
and a disk black body emission {\tt diskbb} (Mitsuda et al.~1984; Makishima et al.~1986). 
We first tried to approximate an absorption or emission line with a gaussian, 
because statistical errors are not small enough 
to determine the precise line profile. 
To make spectral changes more clear, 
we took the ratios of the four spectra to the model adjusted to fit the spectrum of Phase A. 
By looking at the ratios as illustrated in Figure~3b, 
we find it more obvious 
that the spectrum hardens as the flux increases 
and the absorption is deepest in Phase C. 

\subsection{Quantification of the absorption line} 

%
\begin{table*}
  \caption{Best-fit parameters of the XIS0 spectrum for each phase and an entire observation.}
  \label{tab:fit}
  \begin{center}
  \begin{tabular}{llllll} \hline \hline              
  Parameters            & A                & B & C & D &  All       \\ \hline  
  He-like Fe$_{\rm{XXV}}$ absorption line   &               &  &  &  &  \\
  Ec (keV) &  6.698(fix)    &  6.698(fix) & 6.614$^{+0.034}_{-0.085}$ &  6.698(fix) & 6.664$^{+0.025}_{-0.013}$        \\  
  $\sigma$  (eV)   &  0.1~(fix)             &  0.1~(fix) & 90.5$^{+100.1}_{-51.5}$ &  0.1~(fix) & 32.8$^{+32.7}_{-32.8}$        \\
  Norm ($10^{-3}$ ph cm$^{-2}$ s$^{-1}$) &  0.60$^{+0.55}_{-0.40}$             & 2.31$^{+0.89}_{-0.86}$ & 10.3$^{+2.24}_{-1.56}$ & 3.40$^{+1.01}_{-1.00}$ & 2.94$^{+0.49}_{-0.35}$        \\  
  EW (eV) &  11.8$^{+10.8}_{-7.9}$            & 20.4$^{+7.9}_{-7.6}$ & 66.0$^{+14.4}_{-10.0}$ & 24.6$^{+7.3}_{-7.2}$ & 24.3$^{+4.1}_{-2.9}$        \\[0.5em]  
  H-like Fe$_{\rm{XXVI}}$ absorption line$^a$   &               &  &  &  &  \\
  Norm. ($10^{-3}$ ph cm$^{-2}$ s$^{-1}$) &  0$_{-0.13}$  & 0.02$^{+0.01}_{-0.02}$ & 0.58$^{+1.71}_{-0.58}$ & 3.40$^{+0.98}_{-0.98}$ & 0.50$^{+0.50}_{-0.50}$        \\  
  EW (eV) &  0$_{-0.03}$  & 0.2$^{+1.0}_{-0.2}$ & 4.5$^{+13.3}_{-4.5}$ & 9.3$^{+2.7}_{-2.7}$ & 4.8$^{+4.8}_{-4.8}$        \\[1em]  
  Neutral Fe K$_{\alpha}$ narrow emission line$^b$   &               &  &  &  &  \\	
  Norm. ($10^{-3}$ ph cm$^{-2}$ s$^{-1}$) &  0.4$^{+0.4}_{-0.4}$ & 1.2$^{+0.9}_{-0.9}$ & 2.6$^{+2.2}_{-1.9}$ & 1.0$^{+1.1}_{-1.0}$ & 0.9$^{+0.4}_{-0.5}$        \\[0.5em]  
  Neutral Fe K$_{\alpha}$ broad line$^b$   &               &  &  &  &  \\	
  Norm. ($10^{-2}$ ph cm$^{-2}$ s$^{-1}$) &  1.20$^{+0.22}_{-1.20}$  & 0.23$^{+0.25}_{-2.54}$ & 1.12$^{+0.56}_{-1.88}$ & 8.95$^{+2.83}_{-2.60}$  & 5.12$^{+0.22}_{-0.20}$        \\[0.5em]  
  Continuum$^c$   &               &  &  &  &  \\	
  $\tau$$^d$  &   0.21$^{+0.04}_{-0.04} $  &  0.16$^{+0.08}_{-0.06} $  & 0.33$^{+0.05}_{-0.05} $ & 0.06$^{+0.07}_{-0.06} $ & 0.12$^{+0.02}_{-0.02} $        \\
  $N_{\rm{dbb}}$$^e$ (10$^{5} R_{\rm{in}}^{2} D_{\rm{10}}^{-2} \cos\theta$)          & 1.66$^{+0.18}_{-0.21} $              & 1.79$^{+0.96}_{-0.60} $  & 0.72$^{+0.92}_{-0.72} $ & 2.33$^{+1.12}_{-1.44} $ & 1.47$^{+0.15}_{-0.16} $ \\ 
  $\Gamma$          & 2.67$^{+0.06}_{-0.12}$              & 2.36$^{+0.15}_{-0.14} $ & 2.16$^{+0.17}_{-0.16} $  & 2.93$^{+0.27}_{-0.24} $ & 2.71$^{+0.02}_{-0.02}$        \\ 
  $N_{\rm{pl}}$ (ph keV$^{-1}$ cm$^{-2}$ s$^{-1}$ at 1keV)          & 7.96$^{+0.88}_{-1.64}$ & 9.88$^{+4.23}_{-1.30}$ & 8.92$^{+4.30}_{-1.29}$  & 27.5$^{+15.2}_{-8.7}$ & 17.5$^{+2.51}_{-0.10}$        \\[1em] 
  $\chi^{2}_{\nu}$ (d.o.f)           & 1.07(160)      & 0.98(160) & 1.01(158) & 1.12(160) &   1.18(158)  \\[0.5em] \hline
  {\tt Kabs}$^f$ [$b= 100$~km s$^{-1}$]   &               &  &  &  &  \\	
  $z$ (10$^{-3}$) &   0. (fix)  &  0. (fix)  & 9.4$^{+2.5}_{-3.6} $ & 0. (fix) & 3.0$^{+2.5}_{-2.7} $        \\
  $N_{Fe_{\rm{XXV}}}$ (10$^{19}$ cm$^{-2}$) &  0.07$^{+0.24}_{-0.06}$  &  0.29$^{+0.54}_{-0.23}$  & 6.54$^{+3.28}_{-2.65}$ & 0.55$^{+0.62}_{-0.37}$  & 0.59 $^{+0.34}_{-0.25} $        \\
  $\chi^{2}_{\nu}$ (d.o.f)           & 1.07(160)      & 0.98(160) & 1.02(159) & 1.11(160) &   1.18(159)      \\[0.5em]
  \hline \hline 
  
\end{tabular}
\end{center}
\begin{itemize}
\setlength{\parskip}{0cm} %
\setlength{\itemsep}{0cm} %
\item[$^a$] The value of $E_{\rm{c}}$ and the width is fixed at 6.966 keV and 0.1 eV, respectively. 
\item[$^b$] The value of $E_{\rm{c}}$ is fixed at 6.4 keV, while the width at 0.1 eV for the narrow line and 1.0 keV for the broad line. 
\item[$^c$] The neutral column density is fixed at 7.0$\times$10$^{21}$ cm$^{-2}$. 
\item[$^d$] The energy of the edge is fixed at 7.5 keV. 
\item[$^e$] The temperature of the disk is fixed at 0.4 keV. $R_{\rm{in}}$ and $D_{\rm{10}}$ are inner radius and the distance in a unit of 10 kpc. 
\item[$^f$] A {\tt Kabs} model for He-like Fe K$_{\alpha}$ is used in place of a negative gaussian at $\sim$ 6.7 keV. 

\end{itemize}
\end{table*}

We then tried to quantify the four phase-sliced spectra and the entire spectra (Phase~All). 
We first checked whether two absorption lines 
at $\sim$6.7 and $\sim$ 7.0 keV are statistically significant 
by fitting the five spectra over the narrow energy range 
around these lines with {\tt gauss+power}. 
The former is found to be significant with higher than 3 $\sigma$ level, 
but the latter is rather marginal with $\sim$ 1 $\sigma$ level for all spectra. 

Then the fits over 4--8 keV are performed with the baseline continuum,  
two narrow negative gaussian as the absorption lines at $\sim6.7$ keV and $\sim$7.0 keV,  
and a broad and narrow positive gaussian as the emission lines at $\sim6.4$ keV; 
i.e., {\tt tbabs*edge*(diskbb+power+gauss($\times$4))}. 
The central energy $E_{\rm{c}}$ and the width of the absorption line for the $\sim 6.7$ keV line 
are fixed at 6.698 keV for He-like Fe$_{\rm{XXV}}$, and 0.1 eV, respectively. 
They are left free for Phase~C and Phase~All, 
while kept fixed for the others. 
The value of $E_{\rm{c}}$ and the width 
for the $\sim7.0$ keV line are fixed at 6.966 keV for H-like Fe$_{\rm{XXVI}}$ and 0.1 eV, respectively. 
They were not freed due to poor photon statistics.
The obtained fitting results are summarized in Table~1. 
Due to the limited energy range,
some of the parameters of the continua are left fixed as shown in the bottom note of Table~1. 
As a result, all fits became statistically acceptable.

The best-fit model and the spectra of Phase C are presented in Figure~3c as an example, 
as well as the spectral ratio to the model from which 
the absorption lines are purposely removed just to visualize the absorption feature. 
The absorption line of He-like Fe$_{\rm{XXV}}$ is evidently detected,  
while that of H-like Fe$_{\rm{XXVI}}$ can be barely seen. 
The values of $E_{\rm{c}}$ is slightly lower than He-like Fe$_{\rm{XXV}}$ (6698eV at rest). 
The difference is larger than a systematic uncertainty $\sim$ 10 eV for the XIS operated with a window option\footnote{http://heasarc.gsfc.nasa.gov/docs/heasarc/caldb/suzaku/docs/ \\ xis/20111228\_MakepiUpdate\_win14GainTable.pdf}, 
suggesting the possibilities of the contribution from lower ionized species than Fe$_{\rm{XXV}}$ and/or redshifted absorption lines.  
The resultant values of a equivalent width (EW) are plotted in Figure~4.  
As expected, the absorption line is gradually deepening from EW $\sim$ 10 eV to EW $\sim$ 60 eV towards Phase~C, 
and then becoming to EW $\sim$ 20 eV. 

For a detailed analysis of the absorption line, 
we used a Voigt profile implemented in {\tt Xspec} 
as a local model of {\tt Kabs} (Ueda et al. 2004, and see also Kubota et al. 2007) 
which has three fitting parameters of the ion column density $N_{\rm{ion}}$, 
the radial velocity dispersion $b$, and a redshift for bulk motion $z$. 
Since the value of $b$ cannot be well constrained by the line profile, 
we fixed $b$ at 100 km s$^{-1}$.
The value of $z$ is left free only for Phase~C and Phase~All. 
The resultant parameters are summarized in Table~1. 
When we changed $b$ into 300 km s$^{-1}$,  
the values of $N_{\rm{Fe}_{\rm{XXV}}}$ and $z$ are 
obtained as 1.70$^{+2.56}_{-1.05}$ $\times 10^{19}$ cm$^{-2}$ and 8.2$^{+3.3}_{-5.0}$$\times10^{-3}$ for Phase C, and 0.82$^{+0.41}_{-0.27}$ $\times 10^{19}$ cm$^{-2}$ and 3.5$^{+2.5}_{-2.6}$$\times10^{-3}$  for Phase~All, respectively, 
which are slightly different from those in the case of $b=100$ km s$^{-1}$ as shown in Table~1. 
Thus, it is certain that the relative value of the column density increases towards Phase~C. 
When we forced $z$ to fix at 0 and added a negative narrow gaussian, 
the fit became acceptable with $\chi^{2}_{\nu}$ (d.o.f)  = 1.01(160), 
resulting in $E_{\rm{c}}$ = $6539^{+38}_{-44}$ eV and EW = 33.5$\pm$10.7 eV. 
Therefore, the data also implies presence of lower ionized species like Fe$_{\rm{XXI}}$. 



%
   
\begin{figure}
\begin{center}
\vbox{}
\includegraphics[width=0.48\textwidth]{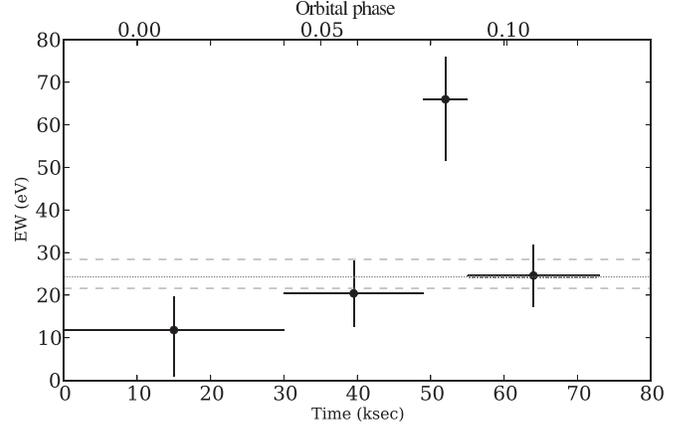}
\caption{
Fitting results of EW for the He-like Fe K$_{\alpha}$ absorption line taken from Phase A to Phase D. 
The mean value and the upper and lower errors taken from Phase~All 
are overlaid in dotted and dashed lines, respectively. 
}
\label{fig:result}
\end{center}
\end{figure}

\section{DISCUSSION AND SUMMARY} 

We have obtained two {\it Suzaku} observations of Cyg X-1 in the high/soft state,  
and found not only the He-like Fe-K$_{\alpha}$ absorption line 
but studied its time evolution from $\phi_{\rm{orb}}$ $\sim$ 0.98 to $\sim$ 0.13. 
Although the absorption line is detected with {\it Chandra} HETG (Feng et al. 2003) 
during the high/soft state, 
it is more obviously detected with {\it Suzaku} XIS 
so that we can map the full evolution from the start and the end of the dip event. 
The detection is clearly obtained at $\phi_{\rm{orb}}$ $\approx$ 0.08, 
which may be slightly later than the {\it Chandra} observation at $\phi_{\rm{orb}}$ $\approx$ 0.
An H-like Fe-K$_{\alpha}$ absorption line appears to be detected, 
though it is not statistically significant.

By dividing the entire span of Obs.~2 into the four phases, 
we obtained spectral evolution as a function of time (Figure.~3ab), 
and quantified their changes (Table~1).
By quantifying the spectra with the conventional models, 
we found that the EW gradually increases from $\sim$ 10 eV ($\phi_{\rm{orb}}$ $\approx$ 0.95) 
to $\sim$ 60 eV ($\phi_{\rm{orb}}$ $\approx$ 0.08), 
and decreases to $\sim$ 20 eV ($\phi_{\rm{orb}}$ $\approx$ 0.15).  
In short, the He-like Fe-K$_{\alpha}$ absorption line increases in EW by a factor of $\sim$ 3 
for less than $\sim$ 10 ks, as shown in Figure~4. 
If absorption lines were driven by the continuous disk wind, 
they could be seen independently of the orbital period; 
e.g., the absorption lines are detected from GRO~J1655-40 
over the entire orbital period without any significant change (Yamaoka et al.~2001). 
Thus, the absorption lines that we found can be 
interpreted as increase in absorbing matter from the stellar wind on that short time scale. 

By fitting the absorption line with the Voigt profile by using the {\tt Kabs} model, 
we evaluated $N_{\rm{Fe}_{\rm{XXV}}}$ and $z$ as summarized in Table~1. 
According to Hanke et al. (2009), 
the value of $N_{\rm{Fe}_{\rm{XXV}}}$ in the low/hard state during the non-dipping period 
is estimated as 1.5$^{+0.3}_{-0.4} \times 10^{17}$ cm$^{-2}$, 
which is consistent with the lower limit that we derived from Phase~A,  $\sim 10^{17}$ cm$^{-2}$. 
The value of $N_{\rm{Fe}_{\rm{XXV}}}$ in Phase C, $\sim 10^{19}$ cm$^{-2}$, 
corresponds to $N_{\rm{H}}$ of $\sim 2 \times 10^{23}$ cm$^{-2}$ when an abundance ratio of Fe to H is 
assumed to be 5 $\times 10^{-5}$ (Anders and Grevesse 1989). 
The redshift of $z$ = 9.4$^{+2.5}_{-3.6} \times 10^{-3}$ obtained from Phase~C 
corresponds to 2900$^{+700}_{-1200}$ km s$^{-1}$ when we take it at its face value.  
The value grossly corresponds to a free fall velocity, 
$\sqrt{GM/r} \sim 4400$ km s$^{-1}$, 
where $G$ is the gravitational constant 
and the distance from the absorber to the source $r$ is 10$^{10}$ cm. 
Alternatively, we can assume that  
an absorbing blob rotating at a distance from the BH to the common center mass, 
$\sim 1.7 \times 10^{12}$ cm (Orosz et al.~2011), 
can traverse our line of sight for $3.4 \times 10^{11}$ cm with a rotational velocity of $\sim 340$ km s$^{-1}$ 
for a duration in Phase C of $\sim 10$ ks.
By putting together all results, 
it can be interpreted 
that some patchy matter provided by the stellar wind, 
presumably located at $r \sim 10^{10-12}$ cm with a hydrogen density 
of $\sim 10^{11-13}$ cm$^{-3}$ or $\sim 10^{(-13)-(-11)}$ g cm$^{-3}$,  
accretes onto and/or transits the BH, 
causing instant decrease in the degree of the ionization of the covering material 
and/or increase in density of the accreting matter. 
Since there could exist several uncertainties on the absolute energy scale of the XIS 
and contamination from lower ionized species indistinguishable from ${\rm{Fe}_{\rm{XXV}}}$, 
continued monitoring with future X-ray calorimeter missions 
such as ASTRO-H SXS (Takahashi et al.~2010) and AXSIO (Bookbinder et al.~2012), 
will allow us to map out the accretion environment more precisely. 


In the low/hard state, 
such a highly ionized Fe-K absorption line has rarely been detected 
though the dips occurs more frequently than in the high/soft state.
It is simply interpreted that 
the absorbing matter in the high/soft state has 
either lower density or higher ionization than in the low/hard state. 
However, 
the number of the photons in energies above $\sim$ 7 keV in the low/hard state 
is more than that in the high/soft state. 
This implies that 
there could be some geometrical effects, 
such as an accretion stream and changes in the scale heights of the X-ray emitting coronae (Wen et al.~1999) or the precession of the accretion disk with an accretion bulge (Poutanen et al.~2008). 

From Obs.~1 performed for $\phi_{\rm{orb}}$ $\approx$ 0.64--0.80, 
we could not see any absorption features (Figure 2). 
Would the appearance of the absorption feature be 
related to the strength of the powerlaw emission 
or we just found the absorption feature when the powerlaw emission is brighter by chance? 
More comprehensive observations for the dipping phenomena 
in different spectral states would 
be crucial to know the whole picture of the stellar wind 
and the accretion flow around the BH.  


\acknowledgements 
The authors would like to express their thanks to {\it Suzaku} team members.
The research presented in this paper has been financed by
the Special Postdoctoral Researchers Program in RIKEN
and JSPS KAKENHI Grant Number 24740129.
S.T. and H.N are supported by Grant-in-Aid for JSPS Fellows.

\bibliographystyle{apj}

\end{document}